\documentclass[preprint]{revtex4-1} 

\usepackage{amsmath}  
\usepackage{amsfonts} 
\usepackage{graphicx} 
\usepackage{multirow}
\usepackage{color}
\usepackage[table]{xcolor}
\usepackage{scrextend}
\usepackage[vskip=0pt]{quoting}

\begin{document}

\title{Review of ``Just Plain Wrong: The Dalliance of Quantum Theory with the Defiance of Bell's Inequality''}

\author{W.M. Stuckey}
\email{stuckeym@etown.edu}
\affiliation{Department of Physics, Elizabethtown College, Elizabethtown, PA 17022}
\author{Timothy McDevitt}
\affiliation{Department of Mathematical Sciences, Elizabethtown College, Elizabethtown, PA 17022}
\author{Michael Silberstein}
\affiliation{Department of Philosophy, Elizabethtown College, Elizabethtown, PA 17022}

\affiliation{}
\author{}

\begin{abstract}
\noindent This is a review of Frank Lad's book ``Just Plain Wrong: The Dalliance of Quantum Theory with the Defiance of Bell's Inequality'' (Austin Macauley Publishers, 2024).
\end{abstract}

\maketitle


\noindent In the Preface of his book, Frank Lad writes \cite[p. 10]{Lad2024}: 
\begin{quote}
    The book you are holding corrects a mistake that … concerns specific influential arguments regarding the defiance of Bell’s inequality that are just plain wrong, embedding a mathematical error that will be understandable to most anyone. The error has deep consequence, leading the community of quantum physicists to reject the principle of local realism, and to proclaim bizarre features of quantum phenomena that the theory does not actually support. Proponents of the error have now been deservedly feted with the award of the Nobel Prize in Physics, 2022, by a Committee who have been taken in by it. Shamelessly, but with due respect, I make bold to call them all to account.
\end{quote}
Yes, Lad claims to have found an error in what ``Literally thousands upon thousands of people'' (p. 202) believe about quantum entanglement and the violation of Bell inequalities (all bare page references are for Lad's book on Kindle \cite{Lad2024}). On pp. 12-13 he writes:
\begin{quote}
    The error occurs in assessing a \textit{gedankenexperiment}, a thought experiment designed to assess the validity of Einstein's insisted principle of local realism. My exposition of the error was first published in the \textit{Journal of Modern Physics} in 2021. [Italics in original.]
\end{quote}
What Lad found are ``functional relations'' in a ``realm matrix'' of hidden variables explaining Bell state entanglement that Lad claims negate what the entire quantum foundations community believes about the violation of Bell inequalities. He accuses the quantum foundations community of committing an ``error of neglect'' (p. 13) by ``neglecting [these] functional relations embedded in a gedankenexperiment'' (p. 14). In order to keep this review from becoming a book itself, we will simply respond to Lad's accusation as he presents it in his \textit{Journal of Modern Physics} paper \cite{lad2021}, ``Quantum Mysteries for No One,'' and in Chapter 4 (p. 193), ``MORE HOOJUMS THAN BOOJUMS:  Quantum Mysteries for No one [sic].'' 

The paper and chapter title refer to ``Quantum Mysteries for Anyone'' \cite{mermin1981a} as reprinted in Mermin's book ``Boojums All the Way through: Communicating Science in a Prosaic Age'' \cite{mermin1991}. Referring to another version of that paper \cite{mermin1981}, Feynman wrote \cite[p. 366-7]{feynman}, ``One of the most beautiful papers in physics that I know of is yours in the American Journal of Physics.'' [Because these two papers \cite{mermin1981a,mermin1981} make the exact same point for the ``general reader,'' we will refer to them collectively as ``Mermin's paper'' hereafter.] Concerning ``hoojums'' Lad writes (p. 195), ``My own allusion now to `hoojums' comes from memory of my mother's usage, referring to `hoojums and boojums' as quasi-mysterious claims that amount to nonsense.'' As we will show, Lad's purported `discovery' does not refute Mermin's introduction to the mystery of quantum entanglement for the ``general reader,'' but actually supports it in an obfuscated fashion. Unfortunately, Lad makes two mistakes regarding the physics of Bell state entanglement which lead him to believe that the results of his ``quantum gedanken simulation'' are only ``reminiscent of the frequency behaviour of [Mermin's] encoded balls'' (p. 229) when in fact his results are totally explicable via Mermin's ``encoded balls.'' These mistakes are the source of Lad's ``bold'' claim that the quantum foundations community is ``just plain wrong'' in the way it analyzes the violation of Bell inequalities.

We will start by reviewing Mermin's result, but before doing so it is important to explain what he did not claim. Mermin's result has nothing to do with experimental tests of Bell's inequality, e.g., ruling out violations of statistical independence regarding hidden variables or ruling out all forms of ``local realism.'' Mermin defines precisely a hypothetical underlying (hidden) physical situation in a reasonable attempt to account for the correlations of spin-entangled particles that, given well-defined assumptions, would produce experimental outcomes that do not agree with the prediction of quantum mechanics (QM) for Bell state entanglement. Whether or not his theoretical situation can be instantiated physically or tested experimentally is absolutely irrelevant to his result. In other words, as long as one does not dispute how to compute and understand measurement outcome probabilities per textbook QM, Mermin's result is a mathematical fact that cannot be refuted. Herein, we assume the standard textbook understanding of QM and note that Lad does not dispute this in his book. With that caveat, let us start with an overview of Mermin's ``Quantum Mysteries for Anyone.'' 

\begin{figure}[h!]
\begin{center}
\includegraphics [height = 50mm]{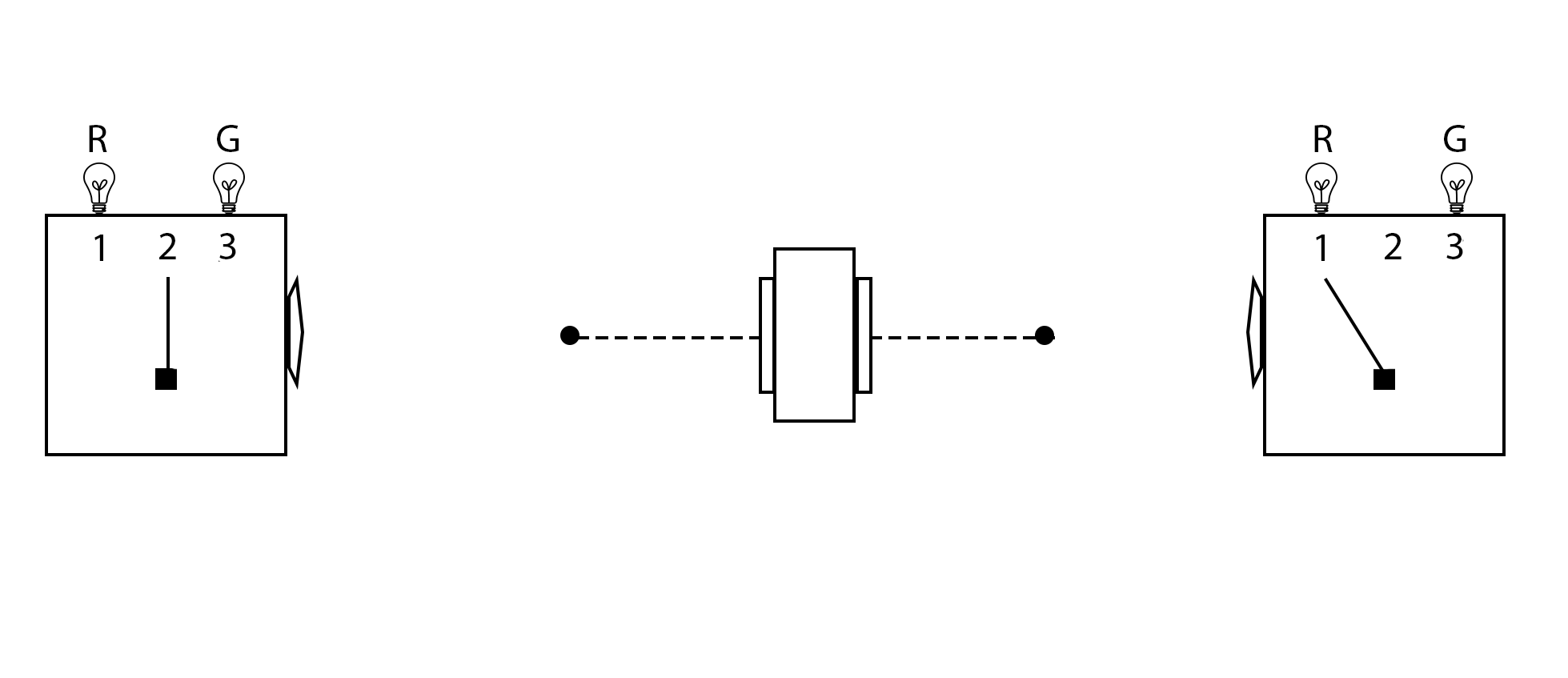}  \caption{\textbf{The Mermin Device}. Alice has her measuring device on the left set to 2 and Bob has his measuring device on the right set to 1. The particles have been emitted by the source in the middle and are en route to the measuring devices. Figure reproduced from Stuckey et al. \cite{MerminChallenge}.} \label{mermin}
\end{center}
\end{figure}

In his paper \cite{mermin1981}, Mermin introduces a device (Figure \ref{mermin}) and explains how it is operated and what it produces. The reader does not need to understand anything about QM to appreciate that this device is indeed mysterious. We will relate the operation of this device immediately to measurements of a Bell spin-$\frac{1}{2}$ singlet state (or ``singlet state'' for short), since this review is written for those with a knowledge of the physics rather than for a ``general reader.'' 

\clearpage

\begin{figure}
\begin{center}
\includegraphics [height = 60mm]{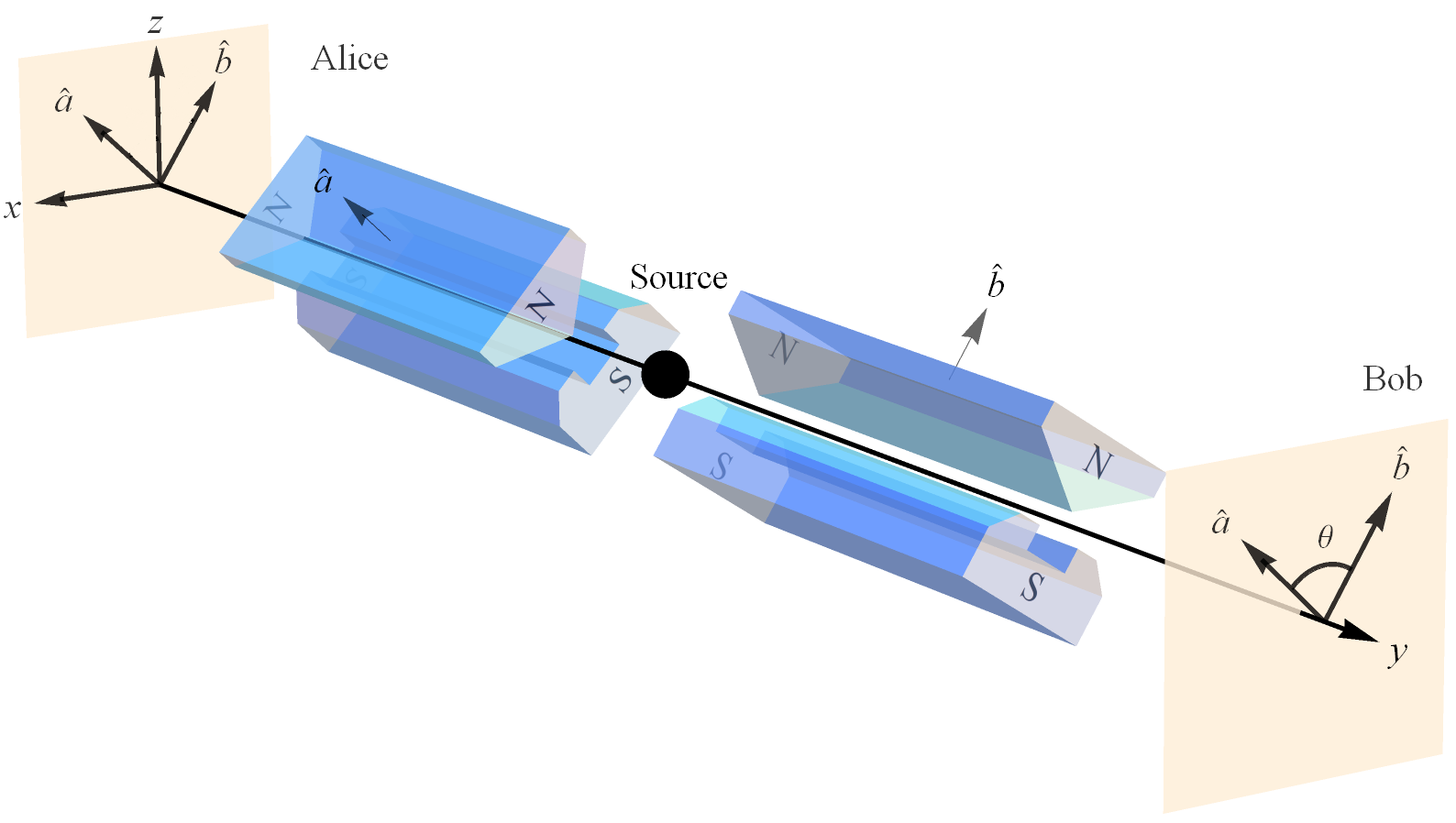}  \caption{Alice and Bob making spin measurements on a pair of particles in a Bell spin-$\frac{1}{2}$ singlet state in the x-z plane perpendicular to the beam line with their SG magnets and detectors. Figure reproduced from Silberstein et al. \cite{silberstein2021}.} \label{EPRBmeasure}
\end{center}
\end{figure}

\begin{figure}[h!]
\begin{center}
\includegraphics [height = 50mm]{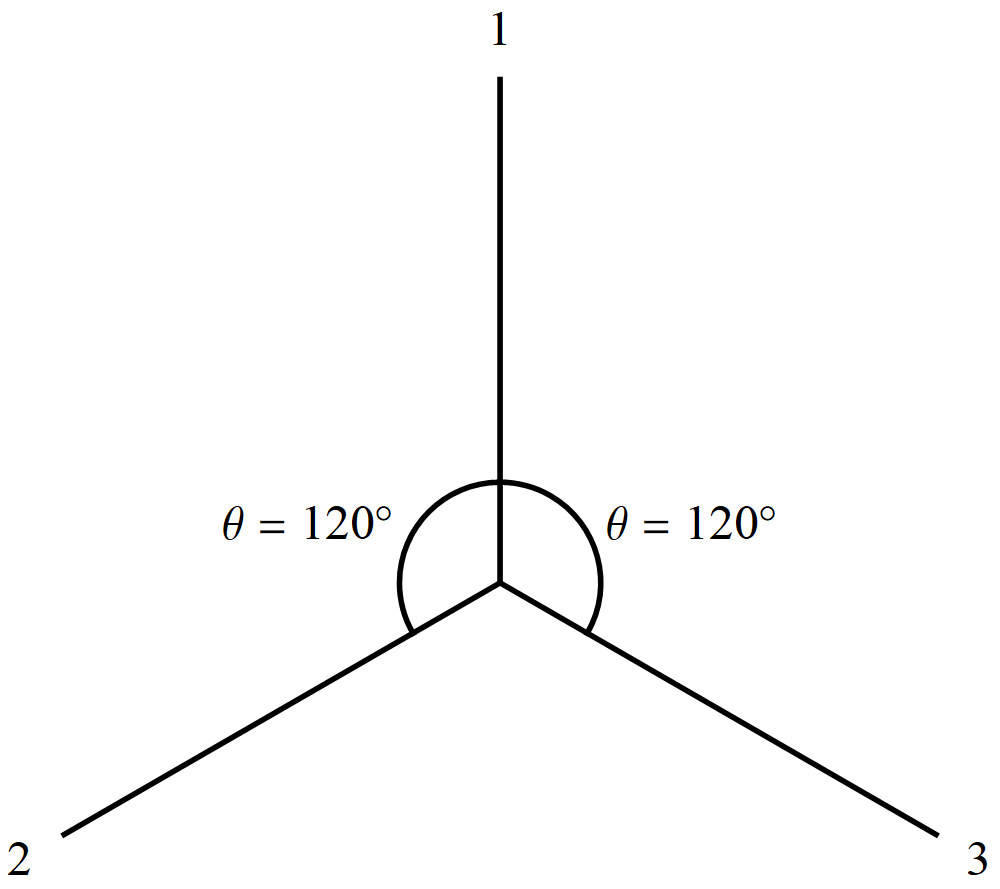} 
\caption{Three possible coplanar orientations of Alice and Bob's SG magnets for a singlet state measurement corresponding to the Mermin device. Figure reproduced from Stuckey et al. \cite{MerminChallenge}.} \label{SGorientations}
\end{center}
\end{figure}

The Mermin device contains a source (middle box in Figure \ref{mermin}) that emits a pair of spin-$\frac{1}{2}$-entangled particles in the singlet state in each trial of the experiment that is measured by Alice and Bob using Stern-Gerlach (SG) magnets (Figure \ref{EPRBmeasure}). The two detectors (boxes on the left and right in Figure \ref{mermin}) controlled by Alice and Bob make measurements at detector settings (1, 2, or 3) corresponding to one of three coplanar SG magnet angles $(0^{\circ},120^{\circ}, \text{or} -120^{\circ})$ (Figure \ref{SGorientations}). The detector settings (1, 2, 3) are selected randomly and independently by Alice and Bob, so we can assume each of the nine detector setting pairs (11, 12, 13, 21, 22, 23, 31, 32, 33) accounts for about $\frac{1}{9}$ of the data. Each measurement at each detector in each trial produces either a result of R (red) or G (green), corresponding to spin up(down) or spin down(up) for Alice(Bob). Here are the two quantum-mechanical facts that produce the ``Quantum Mysteries for Anyone'':
\begin{enumerate}
    \item[Fact 1.] When Alice and Bob's settings happen to be the same in a given trial (``case (a)''), their outcomes are always the same, $\frac{1}{2}$ of the time RR (Alice's outcome is R and Bob's outcome is R) and $\frac{1}{2}$ of the time GG (Alice's outcome is G and Bob's outcome is G).
    \item[Fact 2.] When Alice and Bob's settings happen to be different in a given trial (``case (b)''), the outcomes are the same $\frac{1}{4}$ of the time, $\frac{1}{8}$ RR and $\frac{1}{8}$ GG. 
\end{enumerate}
\clearpage
\noindent Per Fact 1, the Mermin device has totally correlated R-G device outcomes for case (a) corresponding to the totally anti-correlated up-down spin outcomes of the singlet state when Alice and Bob's SG magnets are aligned. Mermin writes \cite{mermin1981}: 
\begin{quote}
Why do the detectors always flash the same colors when the switches are in the same positions? Since the two detectors are unconnected there is no way for one to ``know'' that the switch on the other is set in the same position as its own.
\end{quote}
Thus, Mermin introduces ``instruction sets'' to account for Fact 1 per ``local realism.'' That is, in each trial of the experiment each particle (``encoded ball'' in \cite{mermin1981a}) in the pair carries the same rule or property for producing an outcome in each detector setting (1, 2, or 3). Essentially, we're supposing that the perfect correlations in Fact 1 are explained by the entangled particles possessing the same (hidden) instruction set (GGR, RRG, GRR, RGG, GRG, RGR, GGG, or RRR) to account for outcomes in any of the three joint detector settings (11, 22, 33) in local fashion. Concerning the use of instruction sets to account for Fact 1, Mermin writes \cite{mermin1981a}, ``I cannot prove that it is the only way, but I challenge the reader, given the lack of connections between the detectors, to suggest any other.'' Again, it is important to understand that Mermin is assuming this particular instantiation of local realism, so his result is only applicable to his particular use of instruction sets, as we now detail.

Mermin notes first that the particles cannot ``know'' what detector settings they will encounter until they arrive at the detectors. Second, they cannot communicate their settings and outcomes with each other superluminally. Mermin imposes these constraints by stating there are no connections between the detectors or between the detectors and the source, so there is no information flow -- superluminal, retrocausal, or otherwise -- between elements of the Mermin device, except as instantiated by the exchange of the particles themselves. He also tacitly rules out violations of statistical independence for his instruction sets, i.e., ``superdeterminism'' per Hossenfelder \cite{sabine2020}, by assuming that each instruction set produced is ultimately measured with equal frequency (statistically speaking) in all nine detector setting pairs. Note, there is no restriction on the \textit{distribution} of instruction sets, only that each instruction set (not particle pair) that is produced during the many trials of the experiment is measured equally among the detector setting pairs. Finally, Mermin rules out statistical accidents when he writes \cite{mermin1981a}:
\begin{quote}
    The statistical character of the data should not be a source of concern or suspicion. Blaming the behavior of the device on repeated, systematic, and reproducible accidents, is to offer an explanation even more astonishing than the conundrum it is invoked to dispel. 
\end{quote}

\begin{figure}[h!]
\begin{center}
\includegraphics [width=\textwidth]{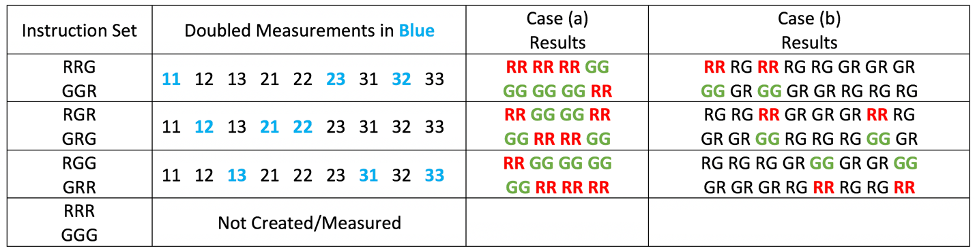}  \caption{Instruction sets can reproduce Facts 1 and 2 for cases (a) and (b), respectively, if they are not measured uniformly in violation of statistical independence. Figure reproduced from Stuckey et al. \cite{NPRF2024}.} \label{merminSD}
\end{center}
\end{figure}

The instruction sets do account for Fact 1, but Mermin shows that they are not compatible with Fact 2, given his assumptions. As Mermin explains, instruction sets with two R(G) and one G(R) will produce agreement in $\frac{1}{3}$ of all case (b) trials. This is where Mermin tacitly rules out superdeterminism. In superdetermism, it's possible that a dynamical mechanism in accord with the initial conditions of the universe causes, for example, Alice and Bob to select detector setting pairs 23 and 32 with twice the frequency of 21, 12, 31, and 13 in those case (b) trials where the source is caused to emit particles with the instruction set RRG or GGR (produced with equal frequency). You can see that this `cosmic conspiracy' would indeed satisfy Fact 2 -- although the detector setting pairs would not occur with equal frequency, but that can be fixed easily as shown in Figure \ref{merminSD}. Then, we will have accounted for Facts 1 and 2 of the Mermin device in accord with local realism with all nine detector setting pairs occurring with equal frequency overall. So, Mermin's assumption that any instruction set that is produced by the source will be measured with equal frequency in all nine setting pairs is necessary to rule out this superdeterministic conspiracy. As we will see, Lad's convoluted analysis of Mermin's instruction sets imposes this constraint de facto and his results conform exactly to Mermin's simple analysis. Let us continue with that analysis. 

 Adding instruction sets RRR and GGG to the distribution then increases this $\frac{1}{3}$ fraction of case (b) agreement. Therefore, the Bell inequality \cite{bell} for the Mermin device is that we expect to get agreement in at least $\frac{1}{3}$ of all case (b) trials. But, Fact 2 for the Mermin device as required by QM says you only get the same outcomes in $\frac{1}{4}$ of all case (b) trials, which violates this Bell inequality. Thus, the mystery of quantum entanglement per the Mermin device is that the instruction sets (with Mermin's assumptions) apparently needed to explain QM's Fact 1 cannot yield QM's Fact 2. Mermin leaves it as a ``challenging exercise to the physicist reader to translate the elementary quantum-mechanical reconciliation of cases (a) and (b) into terms meaningful to a general reader struggling with the dilemma raised by the device'' \cite{mermin1981}. In other words, we know the (elementary) formalism of QM predicts the observed Facts 1 and 2, but we don't have any (consensus) physical mechanism or principle \cite{MerminChallenge,NPRF2024,stuckey2022} to explain or account for that elementary QM formalism. This summarizes the mystery of quantum entanglement and Mermin's challenge as introduced in Mermin's paper.

To remind the reader, ``the elementary quantum-mechanical reconciliation of cases (a) and (b)'' is obtained from the joint probabilities for Alice and Bob's measurements of a singlet state with R and G representing spin up(down) and spin down(up) for Alice(Bob):
\begin{equation}
P(R,R \mid \theta) = P(G,G \mid \theta) = \frac{1}{2} \mbox{cos}^2 \left(\frac{\theta}{2} \right)  \label{QM1jointLike}
\end{equation}
and
\begin{equation}
P(R,G \mid \theta) = P(G,R \mid \theta) = \frac{1}{2} \mbox{sin}^2 \left(\frac{\theta}{2} \right) \label{QM1jointUnlike}
\end{equation}
where $\theta$ is the angle between Alice and Bob's SG detector settings. That is, $\theta = 0^{\circ}$ for case (a), so $P(R,R \mid \theta) = P(G,G \mid \theta) = \frac{1}{2}$ gives Fact 1, and $\theta = \pm 120^{\circ}$ for case (b), so $P(R,R \mid \theta) = P(G,G \mid \theta) = \frac{1}{8}$ gives Fact 2. Obviously, Mermin's device is an accurate representation of the elementary QM, and instruction sets per Mermin's assumptions do not reproduce the QM predictions, so his result is a mathematical fact that cannot be refuted. Thus, we know that Lad must be mistaken when he claims (p. 196), ``I explain in this chapter the serious consequences of his mistaken representation of Bell’s theorem and its implications.''

Lad's first mistake concerns the mystery that the instruction sets required to explain Fact 1 fail to explain Fact 2. He writes (p. 202):
\begin{quote}
    Would you like to dwell on this puzzle yourself for a while if you have not already done so? Knock yourself out. Literally thousands upon thousands of people have done so, and have been taken in by a sleight of hand in the argument. Without warning or fuss, Professor Mermin has switched the setting of the experimental runs on us. Rather than counting the observed sequence of spin-products as each pair of balls [representing the particles emitted by the source] enters the machine at a selected dial setting [detector setting], he is counting the spin-products for each pair of balls as if it would pass all six of the mixed dial settings. His reported lighting statistics pertains to one ballgame, and his counting of the matching colours pertains to another, two completely different games. 
\end{quote}
Despite what ``Literally thousands upon thousands of people'' find quite clear about Mermin's explanation of the experimental procedure, Lad misunderstands what Mermin is saying regarding the conduct of the experiment with instruction sets. He believes Mermin is describing two different experiments. 
So, concerning Mermin's statement \cite{mermin1981a}:
\begin{quote}
Suppose, for example, that both particles carry the instruction set RRG. Then out of the six possible case (b) detector settings, 12 and 21 will result in both detectors flashing the same colour (red), and the remaining four settings 13, 31, 23, and 32, will result in one red flash and one green. Thus, both the detectors will flash the same color for two of the six possible case (b) detector settings.
\end{quote}
Lad writes \cite{LadEmail}:
\begin{quote}
    [Mermin] is proposing very clearly the gedanken procedure of sending a single selected pair of balls [particles] to all six of the mixed dials settings, and counting the numbers of various possible light signals that will result.
\end{quote}
But if Lad's inference is correct, then these statements by Mermin make no sense \cite{mermin1981a}: 
\begin{quote}
    Let us now consider the totality of all case (b) runs. In none of them do we ever learn what the full instruction sets were, since the data reveal only the colors assigned to two of the three settings. (The case (a) runs are even less informative.)
\end{quote}
and
\begin{quote}
    In the case of my device, three such properties are involved for each particle. We will call them the 1-color, 2-color, and 3-color of the particle. The \textit{n}-color of the particle is red if a detector with its switch set to \textit{n} flashes red when the particle arrives. The three \textit{n}-colors of a particle are complementary properties. The switch on a detector can be set to one of only three positions, and the experimental arrangements for measuring the 1-, 2-, or 3-color for a particle are mutually exclusive. (We may assume, to make this point quite firm, that the particle is destroyed by the act of triggering the detector, which is, in fact, the case in many recent experiments probing the principles that underly the device.)
\end{quote}
Clearly, as explained above and confirmed directly by Mermin himself \cite{MerminEmail2022}, there is only one experimental procedure and in it each pair of particles is measured only once in one of the nine detector setting pairs. It appears that Lad is conflating instruction sets with particle pairs. In the many trials of the experiment overall, each instruction set will be measured in all nine detector setting pairs with equal frequency, but each particle pair will only be measured in one detector setting pair. All of this was explained to Lad in email exchanges in 2022, so we do not understand why Lad has persisted in this erroneous accusation. Again, Mermin is simply showing that QM Facts 1 and 2 for this experiment cannot be accounted for by assuming the particles have instruction sets per his assumptions (and this is a mathematical fact). As Mermin explains, while we don't know exactly which instruction set was being measured in each case (b) trial, we do know the Bell inequality for the case (b) trials will be satisfied if the particles possess any distribution of the hidden instruction sets (being measured as described above). Since Fact 2 violates this Bell inequality, we know that the QM prediction is not consistent with Mermin's instruction sets. 

Again, Facts 1 and 2 for the Mermin device are in accord with coplanar SG spin measurements at (0, 120, or -120) degrees for a singlet state, i.e., in accord with QM. In this QM experiment, everyone understands (including Lad) that each particle pair is in the same singlet state and each particle pair is measured randomly in one (and only one) of the nine detector setting pairs (the three of case (a) and the six of case (b)). Contrary to Lad's claim, Mermin is not changing anything about the conduct of that experiment when he introduces instruction sets in an attempt to account for Facts 1 and 2 of said experiment in accord with local realism. Lad (correctly) understands that while there are multiple measurements in the spin experiment, there is only a single measurement of each particle pair. But for some reason when instruction sets are assumed to exist in that experiment, he erroneously believes that each pair of particles (rather than each instruction set) is subsequently measured in all nine setting pairs, always responding as specified by its original instruction set. So, in Lad's local realism model the instruction set for each particle pair survives unaltered from one of the nine measurements to another, which certainly would not happen for electrons subject to the noncommuting SG measurements shown in Figure \ref{SGorientations}. In other words, not only does Lad's local realism model not reproduce Fact 2 for the entangled-spin measurements, but it would also not reproduce his subsequent (totally irrelevant) sequence of non-entangled-spin measurements. 

\begin{figure}[h!]
\begin{center}
\includegraphics [height = 85mm]{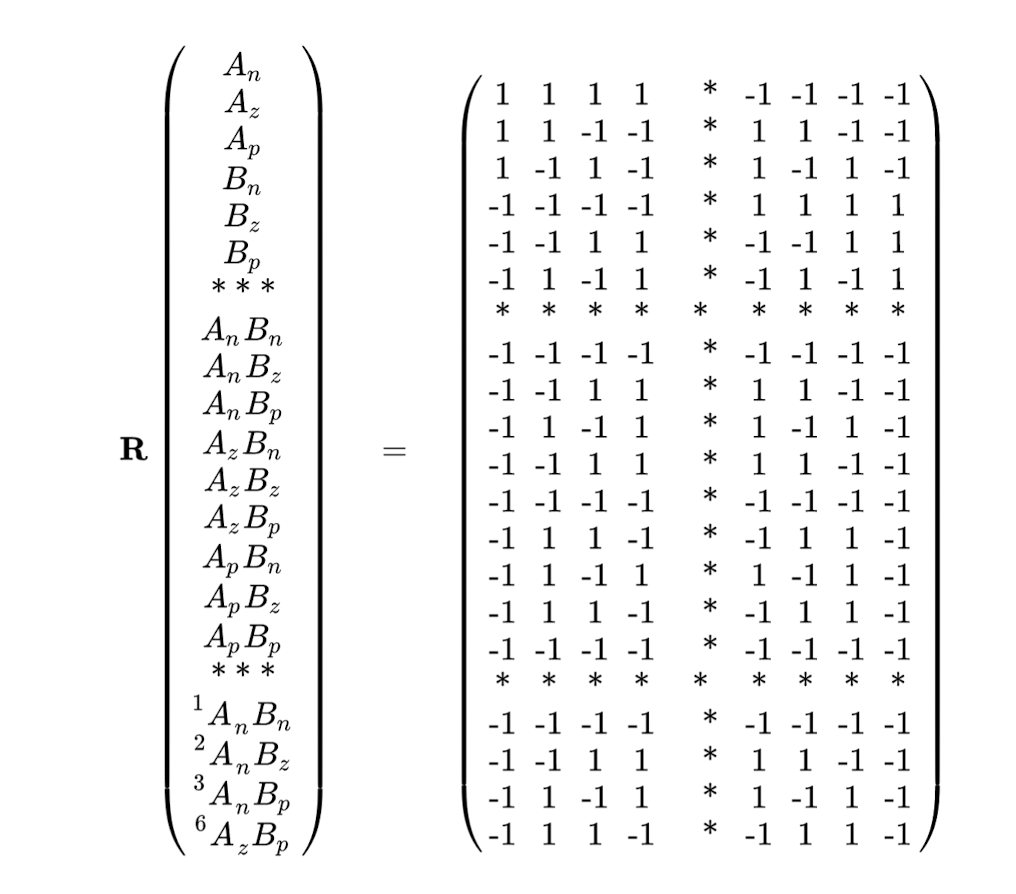} 
\caption{The bottom third of the realm matrix shows the four unique rows of the middle two cells. This will not concern us.} \label{realm}
\end{center}
\end{figure}

Since Lad (mistakenly) believes that each pair of particles emitted in accord with local realism is subsequently measured in all nine setting pairs, he organizes the data for his (not Mermin's) locally real account of the experiment into 9-dimensional vectors of data he calls ``G9 vectors.'' Of course, this produces data that is equivalent to having each instruction set (not each particle pair) measured in all nine settings with \textit{exactly} equal frequency (i.e., in very strict accord with statistical independence), so it cannot produce results at odds with Mermin's analysis of instruction sets. As we will see, contrary to Lad's claim, his labyrinthine version of instruction sets does not produce results that cannot be explained by Mermin's very simple version. Returning to Lad's version, each G9 vector contains the measurement results for each instruction set in all nine detector setting pairs. Each component of each G9 vector is -1 or +1 depending on whether the corresponding instruction set produced the same outcome (-1) or a different outcome (+1) in each detector setting pair (11, 12, 13, 21, 22, 23, 31, 32, 33). Lad writes (p. 205):
\begin{quote}
    Suppose we do send each pair of identical encoded balls to all nine paired dial settings, ordered as 11, 12, 13, 21, 22, 23, 31, 32, 33, and numbered 1 to 9. Designating observations of matching light colours by a -1, and mixed-light-colour observations by +1, the experimental results would need to be recorded not merely by something like 13RG, but rather by something more extensive, such as (-1,+1,+1,+1,-1,-1,+1,-1,-1).
\end{quote}
Lad constructs what he calls the ``realm matrix'' (p. 216) from the instruction sets (that he calls ``G6 vectors'') and the G9 vectors (Figure \ref{realm}). The instruction sets are the top two cells of the realm matrix partition and the G9 vectors are the middle two cells in the realm matrix partition of Figure \ref{realm}. He writes (p. 219):
\begin{quote}
    A \textit{substantive} matter to recognise about this realm matrix is that the four columns of the right partition of the middle section constitute a \textit{reversed replica of the columns of the left partition} of that section. [Italics in original.]
\end{quote}
Of course, that simply reflects the fact that the G9 vector for the instruction set RRG is equivalent to that for GGR, and that for RGR is equivalent to that for GRG, etc. This is not the least bit surprising, but Lad seems to find such structure in the realm matrix important. Again, keep in mind that Lad is simply manipulating Mermin's instruction sets in a particular (tortuous) fashion here. Therefore, the four unique G9 (column) vectors in the realm matrix correspond to the four rows of our Table \ref{tab:G9}; he calls those four unique columns the ``realm matrix R9-4.'' Note that Lad also uses the following notation for the detector setting pairs of his realm matrix: ($A_nB_n, A_nB_z, A_nB_p, A_zB_n, A_zB_z, A_zB_p, A_pB_n, A_pB_z, A_pB_p$). We will stick to (11, 12, 13, 21, 22, 23, 31, 32, 33) as in Table \ref{tab:G9} per the Mermin device, since there is no substantive reason to alter the conventions of the device.

\begin{table}
$$
\begin{array}{|c|ccccccccc|c|}
    \hline
    \mbox{Instruction Sets} & \multicolumn{9}{c|}{\mbox{Setting Pair and G9 Data}} & \mbox{Data Vector}\\
    \hline
    & 11 & 12 & 13 & 21 & 22 & 23 & 31 & 32 & 33 & \\
    \cline{1-11}
    \mbox{GGG \hspace{1mm} RRR} & -1 & -1 & -1 & -1 & -1 & -1 & -1 & -1 & -1 & \mbox{G9-1}\\
    \mbox{GGR \hspace{1mm} RRG} & -1 & -1 & +1 & -1 & -1 & +1 & +1 & +1 & -1 & \mbox{G9-2}\\
    \mbox{GRG \hspace{1mm} RGR} & -1 & +1 & -1 & +1 & -1 & +1 & -1 & +1 & -1 & \mbox{G9-3}\\
    \mbox{GRR \hspace{1mm} RGG} & -1 & +1 & +1 & +1 & -1 & -1 & +1 & -1 & -1  & \mbox{G9-4}\\
    \cline{1-11}
\end{array}
$$
\caption{{\bf Instruction Sets and Their Corresponding G9 Vectors.}}
\label{tab:G9}
\end{table}

This (p. 220) is where Lad points to his discovery of twelve ``functional relations $\{-1,+1\}^2$ into $\{-1,+1\}^7$ that inhere within the structure of the realm matrix R9-4.'' That is, any two rows in R9-4 that contain $\{+1,+1\},\{+1,-1\},\{-1,+1\},\{-1,-1\}$ between them can be considered the domain of a function into the other seven rows and there are twelve such pairs of rows in R9-4. For example, rows 2 and 3 in R9-4 (corresponding to columns 12 and 13 in Table \ref{tab:G9}) are the domain of one such functional relation and Lad describes this as $23 \rightarrow 1456789$. These twelve functional relations constitute the fifteen possible pairings of case (b) detector setting columns in Table \ref{tab:G9} minus the three pairings 12 with 21, 13 with 31, and 23 with 32. To attribute any significance to this is to attribute significance to the way you try to make instruction sets account for Facts 1 and 2 of the Mermin device. And, again, the instruction sets will not account for Fact 2 no matter how you distribute them, as Mermin showed very simply. 

Nonetheless, Lad attributes great significance to the existence of these functional relations in the realm matrix R9-4 and sets about using them in Monte Carlo simulations. Lad generates 1,000,000 G9 vectors in each of his twelve Monte Carlo simulations corresponding to each of the twelve functional relations (1,000,000 G9 vectors correspond to 9,000,000 trials with the Mermin device). Notice that it is possible to find distributions of the four G9 vectors (and therefore, of the instruction sets) so as to satisfy Fact 2 in four of the six case (b) detector setting pairs (Table \ref{tab:G9-1-2-3}). Of course, the total for such distributions is still $\frac{1}{3}$ (Bell inequality), so it fails to account for Fact 2 (must be $\frac{1}{4}$ for every case (b) setting pair per QM) exactly as Mermin explained.

\begin{table}
$$
\begin{array}{|ccccccccc|}
    \hline
    \multicolumn{9}{|c|}{\mbox{Fraction of -1 Results}} \\
    \hline
    11 & 12 & 13 & 21 & 22 & 23 & 31 & 32 & 33\\
    \hline
     1.00 & 0.25 & 0.25 & 0.25 & 1.00 & 0.50 & 0.25 & 0.50 & 1.00\\
    \hline
\end{array}
$$
\caption{{\bf 1:1:2 Distribution Ratio of G9-2:G9-3:G9-4 Data Vectors.}\\ The total fraction of -1 results for case (b) is $\frac{1}{3}$ for these G9 vectors (corresponding to Mermin's two R(G) and one G(R) instruction sets), but there are several case (b) setting pairs with exactly a fraction of $\frac{1}{4}$ in agreement with QM. There are no G9-1 vectors (corresponding to Mermin's RRR and GGG instruction sets) used here. This is entirely consistent with Mermin's analysis, as it must be.}
\label{tab:G9-1-2-3}
\end{table}

\clearpage

\begin{figure}[h!]
\begin{center}
\includegraphics [height = 30mm]{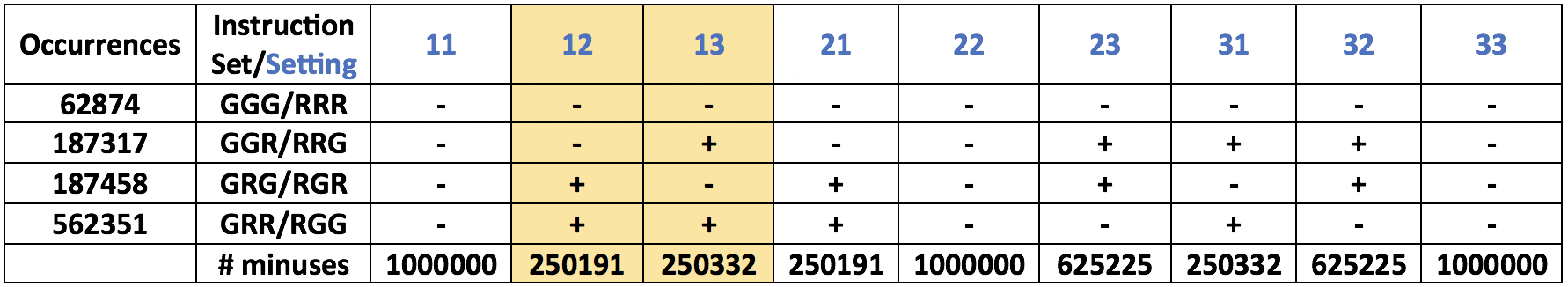} 
\caption{Combining Tables \ref{tab:G9}, \ref{tab:Lad-23}, and \ref{tab:G9Distribution} for the ``23'' functional relation. Each element of the ``\# minuses'' row gives the number of minuses in the column above it.} \label{Results}
\end{center}
\end{figure}

This is where Lad makes his second mistake, he believes Mermin would expect some distribution of the G9 vectors to satisfy Fact 2. That is ridiculous of course, since the G9 vectors simply encode the response of Mermin's instruction sets to all nine detector setting pairs and Mermin shows quite clearly that instruction sets cannot reproduce Fact 2 (given locality and statistical independence). Attributing great significance to the existence of his functional relations and believing (erroneously) that Mermin claims it should be possible for the G9 vectors to generate Fact 2, Lad writes a random number generator to create these $\pm 1$ pairs such that same outcomes in detector settings 12 and 13 occur approximately 25\% of the time for the ``23'' functional relation (similarly for the other eleven functional relations in the other eleven simulations). Of course, each $\pm 1$ pair (element of $\{-1,+1\}^2$) of the domain corresponds to a particular G9 vector (row of Table \ref{tab:G9}), so he can then add up the occurrence of -1 outcomes in any column of his trials to obtain the number of same outcome measurements (-1) for any particular detector setting pair. Table \ref{tab:Lad-23} shows what his Monte Carlo simulation produced for the ``23'' functional relation (the other eleven are very similar, see Lad's Table 2 on his pages 227-8). Figure \ref{Results} depicts the distribution of G9 vectors (Table \ref{tab:G9}) for the 23 $\rightarrow$ 1456789 functional relation (column 23 of Table \ref{tab:G9Distribution}, explained below) that produced the results in Table \ref{tab:Lad-23}.

\begin{table}
$$
\begin{array}{|ccccccccc|}
    \hline
    \multicolumn{9}{|c|}{\mbox{Total Number of -1 Results for 23 $\rightarrow$ 1456789}} \\
    \hline
    11 & 12 & 13 & 21 & 22 & 23 & 31 & 32 & 33\\
    \hline
     1000000 & 250191 & 250332 & 250191 & 1000000 & 625225 & 250332 & 625225 & 1000000\\
    \hline
\end{array}
$$
\caption{This shows the total number of -1 results for each detector setting pair for Lad's Monte Carlo simulation generating 1,000,000 G9 vectors for the ``23'' functional relation.}
\label{tab:Lad-23}
\end{table}

Again, we know from Mermin's very simple and general analysis that any distribution of instruction sets will not reproduce Fact 2. So, we know that Lad's unnecessarily convoluted instantiation of instruction sets via Monte Carlo distributions of G9 vectors related by ``functional relations mapping $\{-1,+1\}^2$ into $\{-1,+1\}^7$'' will satisfy the Bell inequality precisely as explained by Mermin. And indeed, Lad notes that his twelve distributions produce the same outcomes in about $\frac{3}{8}$ (0.375) of the case (b) trials (p. 228). However, on the very next page (p. 229) Lad writes:
\begin{quote}
    These proportions are \textit{not} displayed as three 1's and six 0.25's as proclaimed by Professor Mermin. The three 1's surely appear in the expected places, but of the remaining six columns we find \textit{all} the proportions near 0.375, defying his claim to the proportion arising as $\frac{1}{4}$. [Italics in original.] 
\end{quote}
Why would Mermin (or anyone else) believe this manner of generating distributions of the instruction sets could magically yield Fact 2? Lad himself doesn't seem to realize what he has done. About his 0.375 proportion he writes (p. 229):
\begin{quote}
Indeed, the proportions we have generated in the quantum gedanken simulation are reminiscent of the frequency behaviour of encoded balls exceeding $\frac{1}{3}$.
\end{quote}
and (p. 230):
\begin{quote}
If you do a long sequence of simulated experiments that gedankenly subjects the electrons to all nine paired magnet angle directions in the way local realism restricts them, you would find the proportion of spin-products equal to -1 at about 0.375 whenever the relative angle between magnets equals $-120^{\circ}$ or $+120^{\circ}$. This happenstance governs the counts displayed in [case (b)] columns. \textit{The result has nothing to do with Mermin's proposed explanation of the ``mystery'' involving colour-encoded balls}. [Italics added.]
\end{quote}
For some reason, Lad fails to see that his G9 vector analysis has \textit{everything} ``to do with Mermin's proposed explanation of the `mystery''' using instruction sets. Let us explain Lad's 0.375 proportion of same outcomes for case (b) in terms of Mermin's instruction sets.

\clearpage

You can see that Table \ref{tab:Lad-23} is similar to Table \ref{tab:G9-1-2-3}, so we wanted to know the exact distribution of G9 vectors in each of his twelve Monte Carlo simulations to see how each distribution varies from the 1:1:2 distribution of [G9-2, G9-3, G9-4] shown in Table \ref{tab:G9-1-2-3}. If N1 is the total number of G9-1 occurrences in the simulation for the ``23'' functional relation, N2 is the total number of G9-2 occurrences, N3 is the total number of G9-3 occurrences, and N4 is the total number of G9-4 occurrences, then we can use Table \ref{tab:Lad-23} to generate four equations in N1, N2, N3, and N4:
\begin{center}
N1 + N2 = 250191\\ 
N1 + N3 = 250332\\
N1 + N4 = 625225\\
N1 + N2 + N3 + N4 = 1000000\\
\end{center}
The solutions are given in column 23 of Table \ref{tab:G9Distribution} with the distributions of the G9 vectors for the other eleven functional relations. Instead of 25\%, 25\%, and 50\% of some ordering of G9-2, G9-3, and G9-4 (as in Table \ref{tab:G9-1-2-3}), Lad's simulation is getting (roughly) 19\%, 19\%, and 56\% of some ordering of G9-2, G9-3, and G9-4 while adding 6\% of G9-1, which is responsible for the case (b) agreement increasing from $\frac{1}{3}$ (0.333) to approximately $\frac{3}{8}$ (0.375). All of this is exactly in accord with Mermin's analysis with his easily understood instruction sets per local realism. To see that, referring to Figure \ref{Results}, you can simply subtract out the same outcome contributions for the case (b) detector settings contributed by the G9-1 vector to obtain the number of same case (b) outcomes for G9-2, G9-3, and G9-4 alone:
\begin{center}
Same = 2*(250191 - N1) + 2*(250332 - N1) + 2*(625225 - N1),
\end{center}
count the number of different case (b) outcomes contributed by G9-2, G9-3, and G9-4:
\begin{center}
Different = 4*N2 + 4*N3 + 4*N4,
\end{center}
and take their ratio to show:
\begin{center}
  Different/Same = 2. 
\end{center}

\clearpage

\noindent Using N1, N2, N3, and N4 from column 23 of Table \ref{tab:G9Distribution} in the above equations gives Same = 1874252 and Different = 3748504. This works for all the N1's, N2's, N3's, and N4's in the columns of Table \ref{tab:G9Distribution} using the corresponding rows of Lad's Table 2 to replace the 250191, 250332, and 625225 in the equations above. That is, the number of same outcomes in Lad's case (b) trials counting just the contributions from G9-2, G9-3, and G9-4 vectors gives the $\frac{1}{3}$ (0.333) proportion exactly. Adding the contribution from the G9-1 vector increases that proportion to $\frac{3}{8}$ (0.375). This is exactly in accord with Mermin's use of instruction sets with two R(G) and one G(R) giving $\frac{1}{3}$ (0.333) exactly (regardless of their distribution), and the addition of instruction sets RRR and GGG increasing that proportion to $\frac{3}{8}$ (0.375), as must be the case. 

This highlights a major difference between Lad's use of instruction sets and Mermin's. Because Lad is convinced that the functional relations in his realm matrix R9-4 are physically significant, he must keep the G9-1 vector (corresponding to the RRR and GGG instruction sets) in any distribution or he will destroy the domain of the functional relations. In Mermin's approach, the distribution of instruction sets is totally flexible. But the comparison is moot because neither local realism model agrees with the QM predictions and neither has any empirical support.  

\begin{table}
$$
\begin{array}{|c|cccccccccccc|}
    \hline
    & \multicolumn{12}{c|}{\mbox{G9 Distribution for Each Functional Relation}}\\
    \mbox{Data Vector} & 23 & 26 & 27 & 28 & 34 & 36 & 38 & 46 & 47 & 48 & 67 & 78 \\
    \hline
    \mbox{G9-1} & 62874 & 62527 & 62281 & 62754 & 62756 & 62561 & 62893 & 62410 & 62306 & 62276 & 62595 & 62454\\
    \mbox{G9-2} & 187317 & 187114 & 187815 & 187434 & 188021 & 562898 & 561997 & 187683 & 187334 & 187207 & 562911 & 563037\\
    \mbox{G9-3} & 187458 & 562974 & 187993 & 562506 & 187641 & 187288 & 187726 & 562462 & 187410 & 563382 & 187115 & 187282\\
    \mbox{G9-4} & 562351 & 187385 & 561911 & 187306 & 561582 & 187253 & 187384 & 187445 & 562950 & 187135 & 187379 & 187227\\
    
    \hline
\end{array}
$$
\caption{{\bf Distribution of G9 Vectors for All Twelve Functional Relations in Lad's Simulation.}}
\label{tab:G9Distribution}
\end{table}

The chapter continues for another 26 pages (to p. 256) wherein Lad continues this confusion in his polytope analysis concluding (pp. 251-252):
\begin{quote}
    [Mermin's] claims regarding quantum theoretic prescription of matching lights in $\frac{1}{4}$ of such gedanken observations when the switches differ are blatantly false. ... The probabilities for matching lights proclaimed by Mermin are representable by a nine-tuple vector that does not lie within the convex hull of the coherent vectors supported by quantum theory. We have created a Monte Carlo simulation of results of a scenario that is wholly consistent with this theory and also represents the restrictive symmetric functional relations that derive from the structure of the experiment. It generates proportions of matching lights on the order of 0.375, precisely on the order of magnitude that the professor would have us suspect on account of his shenanigans. 
\end{quote}
Here we see that Lad's conclusion is wrong way round. His Monte Carlo simulation with instruction sets creates a point (0.375,0.375,0.375) inside the polytope while the QM prediction (0.25,0.25,0.25) lies outside (p. 248), so Lad's polytope result obtained using his convoluted G9 vector analysis conforms exactly to Mermin's Bell inequality result from his simple instruction set analysis. Again, we see that Lad erroneously believes his G9 vector analysis ``is wholly consistent with [quantum theory]'' when in fact, his G9 analysis supports the results of instruction sets, which we know from Mermin's straightforward explanation are \textit{not} consistent with QM.

Lad sums up his two mistakes on p. 251 by saying Mermin's claim that spin-products equaling -1 will only occur in $\frac{1}{4}$ of case (b) trials is:
\begin{quoting}
   just plain wrong. Examining the real quantum experiment which the professor would have us ignore [running each particle pair though all nine detector setting pairs], we find embedded within the corresponding thought experiment a surprising feature that has long been unnoticed. Subjecting each pair of electrons to spin-detection at all nine of the paired angle settings would engender an array of restrictive functional relations among the nine observed spin results. The professor neglects these symmetric functional relations mapping $\{-1,+1\}^2 \rightarrow \{-1,+1\}^7$ in his analysis of the gedankenexperiment. His claims regarding quantum theoretic prescription of matching lights in $\frac{1}{4}$ of such gedanken observations when the switches differ are blatantly false.  
\end{quoting}
Again, QM does \textit{not} say Lad's G9 vector results represent what would happen if you actually ran the electrons through nine sequential noncommuting spin measurements. And, Mermin certainly did \textit{not} say the QM prediction of $\frac{1}{4}$ matching lights in case (b) trials represented sequential measurements of the same particle pair. No one with any knowledge of QM would make that claim. Lad was informed of that fact in 2022 by Mermin himself and yet he repeats the claim in his book. Lad writes (p. 10), ``There is shame in denying our mistakes when we are alerted to them, and in actively suppressing their clarification.'' Apparently, that doesn't apply to Lad himself.

In conclusion, Lad's claim that the quantum foundations community has committed an ``error of neglect'' by not using his functional relations to analyze ``the gedankenexperiment'' is itself ``blatantly false.'' ``Literally thousands upon thousands of people'' have correctly used Mermin's instruction sets to derive a Bell inequality that is violated by QM exactly in accord with Lad's G9 vector analysis using his functional relations. Contrary to Lad's belief that he has discovered ``a surprising feature that has long been unnoticed,'' Lad is simply obfuscating Mermin's simple model of local realism, a fact that Lad has ignored since it was pointed out to him in 2022. Overall, it is Lad's understanding of quantum entanglement that is ``just plain wrong.''


\end{document}